\documentclass[aip,pop,preprint,superscriptaddress]{revtex4-1}

\usepackage{graphicx,amsmath,amssymb}

\newcommand{\Igun}{$I_{gun}$}
\newcommand{\Vgun}{$V_{gun}$}

\begin{document}

\preprint{LA-UR-12-25568}

\title{Experimental characterization of railgun-driven
supersonic plasma jets motivated by high energy density physics applications}



\author{S. C. Hsu}
\email[Electronic mail:  ]{scotthsu@lanl.gov}
\affiliation{Physics Division, Los Alamos National Laboratory, Los Alamos, NM 87545, USA}

\author{E. C. Merritt}
\affiliation{Physics Division, Los Alamos National Laboratory, Los Alamos, NM 87545, USA}
\affiliation{University of New Mexico, Albuquerque, NM 87131, USA}

\author{A. L. Moser}
\affiliation{Physics Division, Los Alamos National Laboratory, Los Alamos, NM 87545, USA}

\author{T. J. Awe}
\altaffiliation{Now at Sandia National Laboratories, Albuquerque, NM, USA.}
\affiliation{Physics Division, Los Alamos National Laboratory, Los Alamos, NM 87545, USA}

\author{S. J. E. Brockington}
\affiliation{HyperV Technologies Corp., Chantilly, VA 20151, USA}

\author{J. S. Davis}
\altaffiliation{Now at Applied Physics, University of Michigan, Ann Arbor, MI, USA.}
\affiliation{Physics Division, Los Alamos National Laboratory, Los Alamos, NM 87545, USA}

\author{C. S. Adams}
\affiliation{Physics Division, Los Alamos National Laboratory, Los Alamos, NM 87545, USA}
\affiliation{University of New Mexico, Albuquerque, NM 87131, USA}

\author{A. Case}
\affiliation{HyperV Technologies Corp., Chantilly, VA 20151, USA}

\author{J. T. Cassibry}
\affiliation{Propulsion Research Center, University of Alabama in Huntsville, Huntsville,
AL 35899, USA}

\author{J. P. Dunn}
\affiliation{Physics Division, Los Alamos National Laboratory, Los Alamos, NM 87545, USA}

\author{M. A. Gilmore}
\affiliation{University of New Mexico, Albuquerque, NM 87131, USA}

\author{A. G. Lynn}
\affiliation{University of New Mexico, Albuquerque, NM 87131, USA}

\author{S. J. Messer}
\affiliation{HyperV Technologies Corp., Chantilly, VA 20151, USA}

\author{F. D. Witherspoon}
\affiliation{HyperV Technologies Corp., Chantilly, VA 20151, USA}


\date{\today}

\begin{abstract}

We report experimental results on the parameters,
structure, and evolution of high-Mach-number ($M$)
argon plasma jets formed and launched by a 
pulsed-power-driven railgun.
The nominal initial average jet parameters in the data set analyzed are
density $\approx 2\times 10^{16}$~cm$^{-3}$, electron temperature $\approx 1.4$~eV, 
velocity $\approx 30$~km/s, $M \approx 14$, ionization fraction
$\approx 0.96$, diameter $\approx 5$~cm, and length $\approx 20$~cm.
These values approach the range needed by the Plasma Liner Experiment (PLX),
which is designed to use merging plasma jets
to form imploding spherical plasma liners that can reach peak pressures
of 0.1--1~Mbar at stagnation.
As these jets propagate a distance of approximately 40~cm, 
the average density drops by one order of magnitude,
which is at the very low end of the 8--160 times drop
predicted by ideal hydrodynamic theory of a constant-$M$ jet.


\end{abstract}


\maketitle 

\section{Introduction}
\label{introduction}

\begin{figure}[!h]
\includegraphics[width=2.8truein]{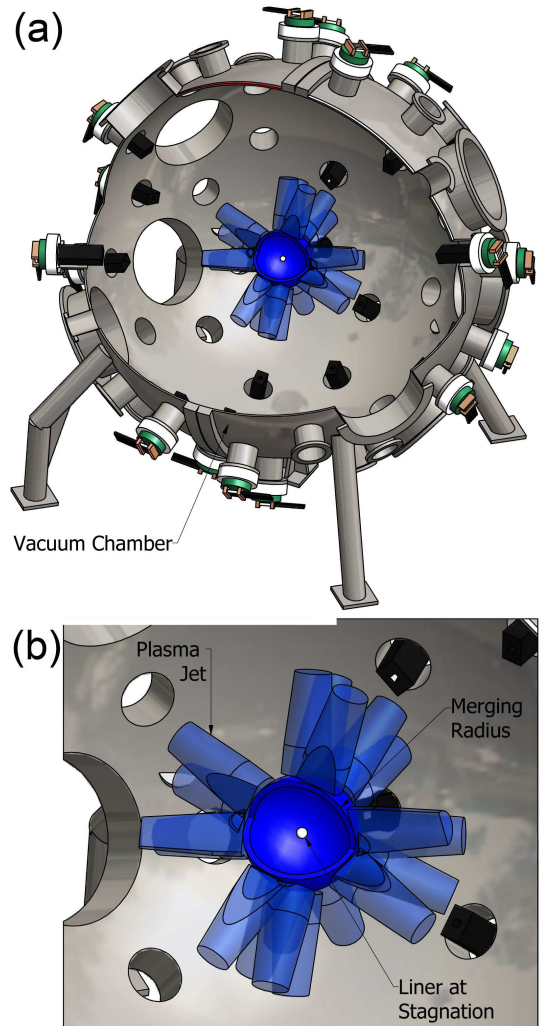}%
\caption{\label{plx-schematic}(a)~Schematic of the Plasma
Liner Experiment (PLX), designed as a
thirty plasma jet experiment to form spherically
imploding plasma liners.  This paper reports single jet characterization
and propagation
studies. (b)~Illustration of  imploding spherical plasma liner formation via the merging
of thirty jets, and also the stagnated liner at peak compression (later in time).}%
\end{figure}

This paper reports 
results from the first in a series of planned/proposed
experiments to demonstrate the formation of imploding
spherical plasma liners via an array of merging 
high-Mach-number ($M$) plasma jets.
Results are obtained on the Plasma Liner Experiment (PLX),
depicted in Fig.~\ref{plx-schematic}, at Los Alamos National Laboratory.
Imploding spherical plasma liners
have been proposed\cite{thio99,thio01,hsu12}
as a standoff compression driver
for magneto-inertial fusion (MIF)\cite{lindemuth83,kirkpatrick95,lindemuth09}
and, in the case of targetless implosions,
for generating cm-, $\mu$s-, and Mbar-scale plasmas for
high energy density (HED) physics\cite{drake10} research.  
Several recent 
theoretical and computational studies
have investigated the physics of imploding spherical plasma liner
formation\cite{parks08,cassibry09,awe11,cassibry12,kim12,davis12,cassibry12b} and
also the fusion
energy gain of plasma liner driven 
MIF\@.\cite{parks08,cassibry09,samulyak10,santarius12,kim12}
In this work, we provide a detailed experimental
characterization of high-$M$ argon plasma jet propagation.
In a forthcoming paper we will present the experimental characterization
of two such jets merging at an oblique angle.  
The next step, a thirty-jet experiment to
form spherically imploding plasma liners that reach 0.1--1~Mbar of peak pressure
at stagnation, has been designed\cite{hsu12,cassibry12b} but not yet fielded.
The primary objective of the single-jet propagation and two-jet oblique merging
studies is to obtain critical
experimental data in order to (1)~uncover important
unforeseen issues
and (2)~provide inputs to and constraints on numerical modeling efforts aimed at
developing
predictive capability for the performance of imploding spherical plasma liners.

This study focuses on addressing issues for
the potential use of railgun-driven high-$M$ plasma jets for 
forming imploding spherical plasma liners and the ability to reach HED-relevant stagnation
pressures ($\gtrsim 1$~Mbar).
Thus, we are interested not only in the jet parameters as they
exit the railgun, but also in the evolution of the plasma jet as it propagates a
distance of $\sim 0.5$~m, as this evolution will affect subsequent jet merging
and ultimately the plasma liner formation and implosion processes.
As such, our work constitutes a unique contribution to the large body
of railgun research,\cite{thio86report}
which has primarily focused on the dynamics and performance
of the ``armature''\cite{thio86,parker89,batteh91} within the railgun bore and the 
ability of railguns to launch solid
projectiles or develop thrust for military\cite{mcnab07} and space
applications.\cite{jahn68,ziemer00,mcnab09}
These jets, if injected into a tokamak plasma, may 
also find applications
in core re-fueling or edge-localized-mode (ELM) pacing.\cite{voronin05nf,liu11nf}

The remainder of the paper is organized as follows:  Sec.~\ref{steps-issues}
gives a brief overview of the physical steps of imploding spherical plasma liner
formation using merging plasma jets, and summarizes key recognized issues (focusing
on jet propagation prior to jet merging); 
Sec.~\ref{setup} describes the experimental setup and diagnostics;
Sec.~\ref{results} presents the
experimental results on plasma jet parameters, structure, and evolution;
and Sec.~\ref{conclusions} provides conclusions and a summary.

\section{Use of plasma jets for forming imploding spherical plasma liners}
\label{steps-issues}


\subsection{Steps of plasma liner formation}
\label{steps}

Here, we provide a brief summary of the steps in imploding plasma liner formation using
an array of merging plasma jets.  Much more detailed accounts,
including both theoretical and computational results, have been presented
elsewhere.\cite{parks08,hsu12,cassibry12b}  

First, multiple plasma jets are launched radially inward from the
periphery of a large spherical vacuum chamber.  The jets propagate separately
until they coalesce at the merging radius $R_m$, which depends on the jet $M$, the number
of jets $N$, the initial jet radius $r_{j0}$, and the chamber radius $R_w$.  For 
non-varying $M$ and the assumption of a jet radial expansion speed of 
$2C_s/(\gamma-1)$,\cite{landau87}
where $C_s$ is the ion 
sound speed and $\gamma$ is the polytropic index, $R_m$ has been derived
as\cite{cassibry12}
\begin{equation}
R_m=\frac{r_{j0}\left(M\frac{\gamma-1}{2}+1\right)+R_w}{1+\frac{2}{N^{1/2}}
\left(M\frac{\gamma-1}{2}+1\right)}.
\label{eq:r_m}
\end{equation}
Note that as $M\rightarrow \infty$ ({\em i.e.}, no radial jet expansion),
$R_m\rightarrow N^{1/2}r_{j0}/2$.  For PLX-relevant
values ($M=14$, $N=30$, $r_{j0}=2.5$~cm, $R_w=111$~cm, $\gamma=1.4$), $R_m\approx 50$~cm,
meaning that the jets will propagate about 60~cm
before merging with adjacent jets.  
The reduced value of $\gamma=1.4$ (below the ideal gas value of $5/3$) is used
throughout the paper to approximate the internal degrees of
freedom of an argon plasma, {\em i.e.}, due to 
ionizations and excitations.\cite{murakami00,awe11}
If $M$ increases during jet propagation
due to radiative cooling, then Eq.~(\ref{eq:r_m}) overestimates $R_m$ and
underestimates the jet propagation distance before merging.  Nevertheless, for
PLX, we are interested in jet propagation distances of order 0.5~m.
Characterizing the evolution of jet parameters over that distance is the focus of this
paper.

Adjacent jets merge at oblique angles at $R_m$
to form an imploding spherical plasma liner.  For PLX, the jet merging
angles are determined
by the vacuum chamber port positions,
with nearest neighbor jets meeting at $\theta=24^\circ$.
Even at this angle, two adjacent jets, each with $M=14$, meet with a relative
Mach number of $2M\sin(\theta/2) \approx 6$.  Thus, shock formation and associated
heating may be expected to contribute significant non-uniformities
to the plasma liner formation process.  The strength of the shock,
and indeed whether shocks even form, are open research questions due to the
parameter regime of our jets,\cite{hsu12,ryutov12} {\em i.e.}, the plasma within
each jet is highly collisional (argon ion mean free path $\lambda_{i}
\approx 4\times 10^{-4}$~cm compared to the jet diameter of 5--20~cm,
where $n_i=2\times 10^{16}$~cm$^{-3}$, $T_i=1.4$~eV, and 
mean charge state $Z_{eff}=1$ have been used to estimate $\lambda_{i}$)
but the interaction between two jets is
semi-collisional or even collisionless (due to the 
high relative velocity between the jets) on the scale of the jet diameter.
Shock heating, should it occur, may degrade the
implosion Mach number, which would lead to the undesirable result of
lower liner stagnation pressure.\cite{awe11,davis12} 
The non-uniformities introduced by discrete jet merging may lead to
asymmetries in the plasma liner implosion that significantly degrade
the peak achievable stagnation pressures.  These and other issues
relating to plasma liner formation via discrete merging plasma jets have
been studied theoretically\cite{parks08} and computationally\cite{cassibry12}
elsewhere and are beyond the scope of this paper. 
Experimental results from PLX on
two-jet oblique merging will be presented in a forthcoming paper.

Finally, after all the jets merge into an imploding spherical plasma liner,
the liner converges with mass density $\rho\sim \rho_0 R^{-2}$,
where $\rho_0$ is the mass density at $R_m$ and $R$ is the radial position
of the imploding liner.\cite{parks08,samulyak10,awe11}
This results in the convergent
amplification of liner ram pressure until the liner reaches the origin
and stagnates (for targetless liner implosions),
converting the liner kinetic energy into stagnation thermal energy
and excitation/ionization energy of the liner plasma.\cite{davis12}  An 
outward-propagating shock is launched, and when this shock meets the trailing
edge of the incoming liner, the entire system disassembles after a stagnation
time $\tau_{stag}\sim L_{j0}/V_{j0}$,\cite{awe11,davis12}
where $L_{j0}$ and $V_{j0}$ are the initial jet length and velocity, respectively.

\subsection{Issues relating to jet propagation}
\label{issues}

There are several issues relating to jet propagation that are important for determining
the performance of the subsequent liner formation, implosion, and stagnation.
Assessing these issues experimentally
is the primary motivation for this single-jet study.  In all cases,
the experimental results are intended to improve the predictive capability
of imploding spherical plasma liner modeling.

The first issue is the achievement of the requisite jet parameters for reaching
the desired liner stagnation pressure.  The design goal for PLX is to reach
0.1--1~Mbar with a total liner kinetic energy of about 375~kJ\@.
A 3D ideal hydrodynamic simulation study\cite{cassibry12b} exploring a wide parameter
space in plasma jet initial conditions
contributed to the PLX reference design calling for thirty argon plasma
jets with initial density $\approx 10^{17}$~cm$^{-3}$, velocity $\approx 50$~km/s,
and mass $\approx 8$~mg.  The simultaneous achievement of these parameters
has recently been demonstrated at HyperV Technologies, to be reported
elsewhere.  For this paper,
we operated at reduced values to extend the lifetime of the railgun
and maximize the number of shots.
The second issue is the evolution of jet parameters, especially
density and velocity, during propagation.  
Developing an accurate predictive capability for modeling plasma liner formation
via merging jets requires accurate knowledge of these jet parameters, both at
the exit of the railgun as well as at $R_m$.
The jet velocity is expected to remain nearly constant, but
density decay arises from jet radial and axial expansion.  
The third issue is the nature of the jet radial and axial profiles because
they are important for accurate modeling of the jet merging and liner formation/implosion
processes.

Experimental values (presented in Sec.~\ref{results}) allow the jet
expansion to be estimated.
In a purely hydrodynamic treatment in which the jet has a non-varying $M$,
both the radial and axial expansion speeds of the jet can be estimated to be
between $C_s$ (jet bulk) and
$2C_s/(\gamma-1)$ (jet edges).\cite{landau87}
For argon at $T_e=1.4$~eV and $\gamma=1.4$, these values are
2.2~km/s and 11~km/s, respectively.  
If the jet travels at 30~km/s over a distance of 50~cm,
this gives a transit time of 16.7~$\mu$s, during which the jet radius will increase by
3.7--16.7~cm and the jet length by 7.4--33.4~cm.  
For a jet with $r_{j0}=2.5$~cm and $L_{j0}=20$~cm,
the jet volume will increase by a factor of 8.4--157.5, and the density will drop
by the same factor.
Due to the nearly factor of twenty
uncertainty in the theoretically predicted density decay, and the fact
that jet cooling 
due to expansion and radiative losses (and thus a varying rate of expansion)
are not accounted for
in the above treatment, it is imperative to determine the
density decay by direct experimental measurement.

\section{Experimental setup}
\label{setup}

\subsection{Plasma Liner Experiment (PLX)}
\label{plx}

Experiments on PLX are conducted in
a 9~ft.\ (2.74~m) diameter stainless steel spherical vacuum chamber,
as depicted in Fig.~\ref{plx-schematic}(a), situated
in a 3000~ft.$^2$ (279~m$^2$) high-bay space with a ten-ton overhead crane.
The vacuum chamber has 60 smaller ports (11~in.\ outer and $7\frac{3}{4}$~in.\ inner
diameters) and 10 larger ports (29.5~in.\ outer and 23.5~in.\ inner diameters);
all flanges are aluminum.
We presently have two operational plasma railguns (see Sec.~\ref{railgun} for
railgun specifications) installed on the vacuum chamber to study
high-$M$ single-jet
propagation, two-jet oblique merging, and two-jet head-on
merging.  This paper reports only single
jet results.
The vacuum base pressure is typically in the low-$10^{-6}$~Torr 
range, achieved with a turbo-molecular pump (3200~l/s Leybold Mag W 3200C, 
$12\frac{3}{4}$~in.\
pumping diameter) backed by an oil-free mechanical pump (Edwards IQDP 80).
During each experimental shot, the turbo-pump is isolated by closing
a gate valve, and thus the vacuum pressure is in the 5--30~$\mu$Torr
range when a plasma jet is fired into the chamber.  The pressure
in the chamber after a shot is in the 0.3-to-few mTorr range;
the gate valve is opened and the chamber evacuated prior
to the next shot.  All chamber pressures are
recorded using an MKS 972B DualMag transducer.
A LabVIEW-based (\texttt{http://www.ni.com/labview}) 
40~MHz field programmable gate array (FPGA) system
controls safety interlocks, shot sequence
including bank charging and dumping, and all trigger signals.
Sensitive control and data acquisition
electronics reside within an electromagnetically shielded cage or
other shielded racks.  All time-series data shown in this paper
were digitized at 40~MHz sampling rate, 12-bit dynamic range, and 1~mV bit resolution
(by Joerger model TR
digitizers).  All times reported in the paper are relative to the trigger time of the
gun rails.  Experimental data and shot data are immediately stored
into an MDSplus (\texttt{http://www.mdsplus.org}) database after every shot.

\subsection{Plasma railgun}
\label{railgun}

\begin{figure}[!h]
\includegraphics[width=3truein]{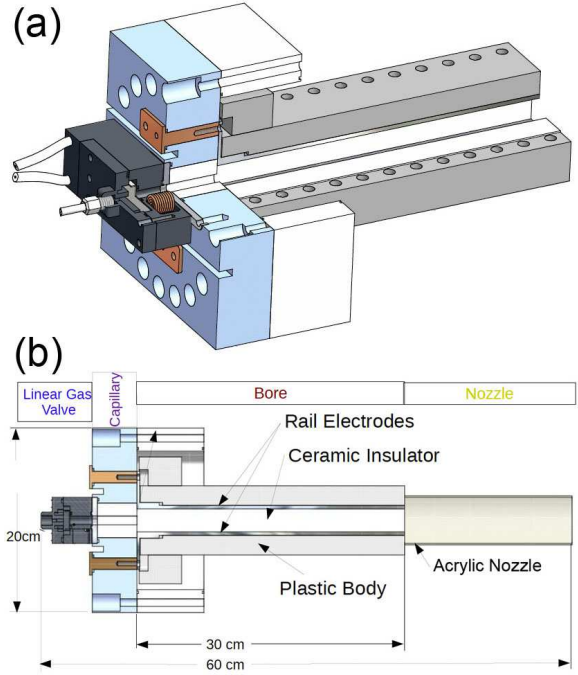}
\caption{\label{gun-schematics}(a)~Three-dimensional
view of the railgun used in this work.  (b)~Side-view schematic of railgun showing
(from left to
right):  fast gas valve, capillary (pre-ionizer), bore (HD-17 tungsten alloy
rails with zirconium-strengthened-alumina insulators), and acrylic nozzle.
The distance from the back of the rails to the end of the nozzle is 47~cm.}
\end{figure}

To achieve the design objectives of PLX ({\em i.e.}, 0.1--1~Mbar of peak
liner stagnation pressure using 
thirty plasma jets with total implosion
kinetic energy of $\sim 375$~kJ) within budgetary constraints,
two-stage parallel-plate railguns
(see Fig.~\ref{gun-schematics}), designed and
fabricated by HyperV Technologies Corp.,\cite{witherspoon11}
were selected as the plasma gun source for
forming and launching 
jets with the requisite parameters (jet density $\approx 10^{17}$~cm$^{-3}$, 
velocity $\approx 50$~km/s, and mass $\approx 8$~mg).
The development, optimization, and performance scaling results of these
PLX railguns,
as well as the simultaneous experimental achievement of the 
aforementioned jet parameters,
will be reported elsewhere. 
Note that
larger coaxial guns with shaped electrodes\cite{witherspoon09,messer09,case10}
are also being developed due to their
suitability for very high current ($>1$~MA) and high jet velocity ($>100$~km/s)
operation, and for having attributes that minimize impurities,
as potentially required for the MIF application.  

\begin{figure}[!htb]
\includegraphics[width=3truein]{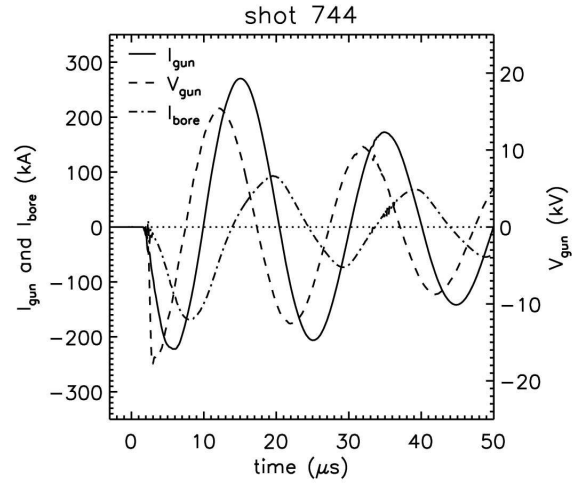}%
\caption{\label{i-v}Representative railgun \Igun, \Vgun, and gun bore
current $I_{bore}$ (near the rear of the rails) for the shots analyzed in this paper.}
\end{figure}

The railgun bore cross-sectional area is
$2.54 \times 2.54$~cm$^2$.
The gun has a fast gas-puff valve (GV), a pre-ionizer (PI), HD-17 (tungsten
alloy) rails housed in a Noryl (blend of polyphenylene oxide and
polystyrene) clamshell body, zirconium toughened alumina (ZTA) insulators between
the rails, and a cylindrical acrylic nozzle (5~cm diameter and 19~cm length).
The results reported here
are from shots with an underdamped, ringing, and slowly decaying gun current
(Fig.~\ref{i-v})
that produces multiple jet structures, with the leading structure having a length of
order 20~cm.  We use ultra-high-purity argon (typically at 18--20~psig) 
for the gas injection, injecting an average total mass of about 35~mg in each shot
for the data set presented in this paper (but only a small fraction of the total
injected mass is contained in the leading jet structure).

The railgun firing sequence is GV followed by the PI and finally
the rails.  The time delays are all adjustable and are chosen for
optimizing certain aspects of jet performance such as density, velocity, or mass.
For the shots reported here,
the GV is fired 300~$\mu$s before the gun rails so that
neutral argon fills the PI volume and the very rear of the railgun bore.
The PI is fired 30~$\mu$s before
the gun rails to break down the neutral gas, giving the PI plasma just enough time
to fill the very rear of the railgun bore.  Then the rails are fired to accelerate the
PI plasma down the bore.  
Control over the plasma jet density and mass 
is mainly through the feed line pressure, GV bank voltage, and 
timing of the neutral gas puff; control
over the jet velocity is mainly through the railgun current and
charge voltage.  
For single-gun operation, the gun, PI, and GV are driven by
36~$\mu$F, 6~$\mu$F, and 24~$\mu$F capacitor banks, respectively,
charged typically to -24~kV, 20~kV, and 8~kV, respectively.
The 36 and 6~$\mu$F banks use 60-kV, 6-$\mu$F Maxwell model 32184 capacitors, and
the 24-$\mu$F bank uses a single 50-kV Maxwell model 32567 capacitor.
All banks are switched by spark-gap switches triggered via optical fibers.
Details about the gun design, operation, performance
scaling, and best achieved parameters
will be reported elsewhere.

We have evidence that the plasma jet is not 100\% argon.
We observe an approximately 25\% higher pressure rise in the chamber
(for the data set presented in this paper)
when the GV and railgun are both fired, compared to when only the GV is fired.
The pressure discrepancy is most likely explained by the railgun current ablating
material from the HD-17 rails and possibly also the ZTA insulators.
We also observe hydrogen, oxygen, and aluminum impurity spectral lines, 
and potentially others that have not yet been identified (with tungsten, nickel, iron,
and copper from the rails as likely candidates).
In the remainder of the paper, we have assumed that the
plasma jets are 100\% argon for the purpose
of interpreting the experimental data because the error introduced is 
generally small compared to diagnostic measurement uncertainties. 
Impurity control in plasma jets is clearly an important issue for the MIF
standoff driver application and requires further study.

\subsection{Diagnostics}

Plasma jet
diagnostics include an eight-chord
interferometer, a visible and near-infrared (IR) survey
spectrometer, an array of three photodiode detectors,
and an intensified CCD (charge-coupled device) imaging camera.
Details of each plasma jet diagnostic system are given below.
Figure~\ref{exp-setup} shows the diagnostic views
in relation to the plasma jet propagation path.
A discussion of the entire planned
PLX diagnostic suite is described in more detail elsewhere.\cite{lynn10}
We emphasize that all plasma jet diagnostic measurements are
averaged quantities over their viewing-chords, and therefore we report
mostly sight-line-averaged quantities.  

Diagnostics for the railgun and pulsed-power systems include 
Rogowski coils (for monitoring railgun, GV, and PI discharge currents),
Pearson current monitors (model 2877) on parallel resistors (for monitoring
instantaneous capacitor bank voltages), and five magnetic probe coils along the railgun
bore (for monitoring electrical current propagation down the bore).
Figure~\ref{i-v} shows representative gun current \Igun, gun voltage \Vgun,
and the gun bore current $I_{bore}$ from the rearmost gun bore magnetic probe.

\subsubsection{Eight-chord interferometer}

An eight-chord fiber-coupled interferometer was designed and
constructed for the PLX project.\cite{merritt12a}
The system uses a 561~nm diode-pumped,
solid-state, 320~mW laser (Oxxius 561-300-COL-PP-LAS-01079) with a
long coherence length ($>10$~m).
Along with the use of single-mode fibers (Thorlabs 460HP) to transport
the laser beams to and from the vacuum chamber, this allows for the use
of one reference chord for all eight probe chords, as long as length mismatches
between reference and probe chords are much smaller than the coherence length.
The fiber-coupling allows for relatively simple chord re-arrangements at 
launch and reception optical bread boards mounted on the vacuum chamber.  The 
beam-splitting and combination optics on the main optical table do not need
to be altered to re-arrange chords.  Large borosilicate windows are used on
both the launch and reception chamber ports.  For this paper, the eight 
laser probe beams (3~mm diameter at the plasma jet)
were arranged transversely to the direction of jet propagation at different
distances from the railgun nozzle: $Z=35.0$--79.5~cm at equal intervals of 
approximately 6.35~cm (see Fig.~\ref{exp-setup}).

The Bragg cell for the interferometer uses a radio frequency
generator (IntraAction ME-1002)
that produces a 110~MHz signal.  The output
signal from the final photo-receivers are passed through bandpass filters
(Lark Engineering MC110-55-6AA) at $110 \pm 55$~MHz to filter out higher-frequency
components of the heterodyne mixing and lower-frequency electrical noise.
The filtered signal is decomposed into two signals, I and Q, proportional
to the sine and cosine of the signal, respectively.  The I and Q signals
pass through a 40~MHz low-pass filter and then are digitally stored (40~MHz,
12-bit resolution, $50~\Omega$ input).
A detailed description of the system
design, components, setup, and electronics are reported elsewhere.\cite{merritt12a}

\begin{figure}[!htb]
\includegraphics[width=3truein]{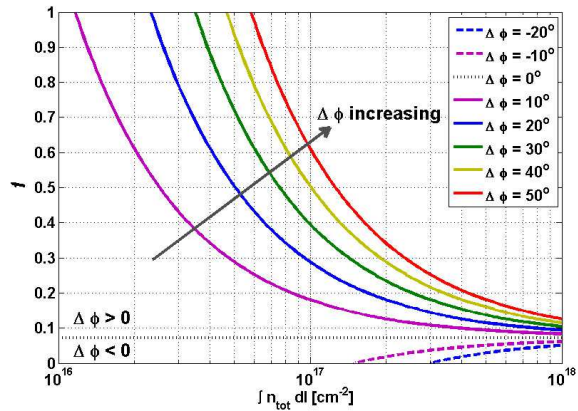}%
\caption{\label{phi-plot}Contours of constant interferometer phase shift 
$\Delta \phi$ 
as a function of ionization fraction $f$ and line-integrated 
argon ion plus neutral density $\int n_{tot}{\rm d}l$,
as calculated using Eq.~(\ref{phase-shift-eq}).}
\end{figure}

For a partially ionized argon plasma, the interferometer phase shift of our
system has been derived as\cite{merritt12b}
\begin{equation}
\underbrace{\Delta \phi}_{[\rm degrees]} = 9.2842 \times 10^{-16} (f-0.07235)
\underbrace{\int n_{tot} {\rm d}l}_{{[\rm cm}^{-2}]},
\label{phase-shift-eq}
\end{equation}
where $f\equiv n_i/n_{tot}$ is the
ionization fraction, $n_{tot}=n_i+n_n$ (where $n_i$ and $n_n$ are
the ion and neutral densities, respectively), and the integral
is over the chord path length.  Note that $\Delta \phi < 0$
when $f<0.07235$.  Figure~\ref{phi-plot} shows selected contours
of constant $\Delta \phi$, calculated using Eq.~(\ref{phase-shift-eq}), as a
function of $f$ and $\int n_{tot}{\rm d}l$.

\subsubsection{Survey spectrometer}

Spectrally resolved plasma self-emission is recorded with a survey spectrometer system
consisting of a 5-mm diameter collimating lens (BK-7),
a 19-element circular-to-linear silica-core
fiber bundle (Fiberguide Industries ``Superguide G,'' 10~m long), a
0.275~m spectrometer (Acton Research Corp.\ SpectraPro 275) with
three selectable gratings (150, 300, and 600 lines/mm), and a gated 1024-pixel
multi-channel-plate array (EG\&G Parc 1420).  All measurements
reported here were taken with the 600~lines/mm grating.   We have taken measurements
from about 300--900~nm, but all the data reported in this paper
are between 430--520~nm.  Two corrections are applied to the raw spectrum for
each shot:  
(1)~attenuation of the fiber bundle (approximately negative linear slope
around -30~dB/km between 430--520~nm) 
and (2)~pixel-dependent factor associated with the position
of the grating in the spectrometer.  The latter correction amplifies
the spectra at low and high pixel-values over the raw spectrum by $\sim 1.7$--5
depending on the exact pixel value.  We placed copper mesh in
front of the collimating lens, as necessary, to keep the peak counts under about
16,000 to avoid saturating the detector.  For the {\em chord} position which
is $Z\approx 41$ from the railgun nozzle
(see Fig.~\ref{exp-setup}), the counts were of order 100, and thus no mesh was used.
Before each shot day, a spectral lamp was used to record line emission
at known wavelengths to provide a pixel-to-wavelength calibration.
The spectral resolution for the data presented in this paper is 0.152~nm/pixel.
The diameter of the viewing chord at the position of the jet,
as imposed by the collimating lens, is approximately
7~cm, which constitutes a nominal 
spatial resolution for the spectroscopy data.
The time resolution is 0.45~$\mu$s
as determined by the exposure (gate) time of the detector.
One spectrum at one time is taken for each shot.

\subsubsection{Photodiode array}

A three-channel photodiode array (PD1, PD2, and PD3)
is used to collect broadband plasma emission for
determining jet propagation speed.  PD1, PD2, and PD3 collect light
mostly transverse
to the direction of jet propagation at $Z=2.7$, 27.7, and 52.7~cm, respectively
(see Fig.~\ref{exp-setup}).  For each channel, 
light is collected through an adjustable aperture positioned in front of
a collimating lens (Thorlabs F230SMA-B, 4.43~mm focal length), which is connected
to a silica fiber that brings the light to the photodiode detectors inside a
shielded enclosure.
The silicon photodiodes (Thorlabs PDA36A) are amplified with variable gain,
and have a wavelength range of approximately 300--850~nm.  The peak responsivity
is 0.65~A/W at 970~nm.  The quoted frequency response decreases with increasing gain.
For the data reported in this paper, the channels at $Z=2.7$, 27.7, and 52.7~cm
had gain settings of 20, 50, and 50~dB, respectively, corresponding
to quoted bandwidths of 2.1, 0.1, and 0.1~MHz.  However, we note
that the observed rise times for the second and third channels are
much faster than the quoted 0.1~MHz bandwidth would dictate
(see Sec.~\ref{sec:velocity}).  All three channels had aperture
openings of $<1$~cm, which constitutes a nominal 
spatial resolution for the photodiode data.  The continuous (in time)
photodiode signals are digitized at 40~MHz by the Joerger TR\@.

\subsubsection{Fast-framing CCD camera}

An intensified CCD camera (DiCam Pro ICCD),
with spectral sensitivity from the UV to near-IR,
is used to capture visible images of the plasma jet.
The camera records $1280\times 1024$ pixel images with 12-bit dynamic range.
The camera can record up to two images per shot. 
However, for inter-frame times below several tens of $\mu$s, the second image often has
``ghosting'' from the first frame.  Thus, in this paper, we recorded only one
frame per shot.  The exposure (gate) is 20~ns.  The camera is
housed inside a metal shielding box and mounted next to a large rectangular
borosilicate window on the vacuum chamber.  A zoom lens (Sigma 70--300~mm 1:4--5.6)
was used for recording the images shown in this paper.
The camera is triggered remotely via an optical fiber.  
To infer quantitative spatial information from the CCD images, we recorded images
with a meter stick held to the end of the railgun nozzle.  Using these calibration
images, we are able to determine pixel-to-centimeter conversion formulae
(see Sec.~\ref{diameter}).  The CCD images presented in this paper
are shown on a logarithmic scale in false color.

\section{Experimental results}
\label{results}

In this section, we present experimental results on the jet parameters 
(including velocity, density, temperature, and ionization fraction), structure,
and evolution for the experimental data set spanning PLX shots 737--819.

\subsection{Jet parameters}

\subsubsection{Velocity}
\label{sec:velocity}

The plasma jet velocity $V_{jet}$
is determined from photodiode array data and corroborated
with interferometer data.  
The velocity is calculated by dividing
the distance between viewing chords by the difference in arrival times of the
peak signal.  We use the arrival time of the peak signal rather than the leading edge
to obtain a more robust estimate of the bulk jet velocity.

Figure~\ref{pd-data} shows the three photodiode signals for a
representative shot (744) from the data set analyzed in this paper.  The signals
are collected along viewing chords intersecting the jet propagation axis at distances
$Z=2.7$, 27.7, and 52.7~cm, respectively, from the end of the gun nozzle
(see Fig.~\ref{exp-setup}).  For this
shot, the peak arrival times are 22.9, 31.6, and 39.8~$\mu$s, corresponding to
average velocities of $V_{12}=28.7$ and $V_{23}=30.5$~km/s between the first and second
pairs of photodiode viewing chords, respectively.
For the entire data set considered in this paper,
$V_{12}=28.9 \pm 3.9$~km/s and
$V_{23}=29.4 \pm 4.5$~km/s, where the uncertainty is
the standard deviation over the data set.
The velocities did not vary significantly across this data set
due to the relatively narrow range of peak gun current (255--285~kA) and
total mass injected into the chamber (pressure rise 1.35--1.75~mTorr,
corresponding to 31--40~mg of argon).  
A similar analysis using the interferometry data (see Fig.~\ref{744-int}(a)
for an example) from the chords at $Z=35.0$ and 47.7~cm
yields an average jet velocity of $34.8 \pm 5.6$~km/s (for the entire
data set), which
compares well with $V_{23}=29.4\pm 4.5$~km/s given above.
The jet velocity appears to be nearly constant 
as it propagates over a nearly 50~cm distance.
At 30~km/s, the jet $M=14$ (assuming that the jet is 100\% argon, $\gamma=1.4$, 
$Z_{eff}=1$, and $T_e=1.4$~eV).

\subsubsection{Density, temperature, and ionization fraction}
\label{n-T-f}

Plasma jet density $n$, electron temperature $T_e$, and ionization
fraction $f$ are determined via a combination of experimental
measurements and interpretation with the aid of theoretical analysis
and atomic modeling.  From spectroscopy, we determine electron density
$n_e$ via Stark broadening of the impurity hydrogen H$_\beta$ line (486.1~nm), 
and we estimate $T_e$ by comparing measured and calculated
non-local-thermodynamic-equilibrium (non-LTE) argon spectra.
Steady-state, collisional-radiation calculations with single-temperature
Maxwellian electron distributions
in the optically thin limit were performed using the PrismSPECT code with
DCA (direct configuration accounting).\cite{macfarlane03}
The following processes were included:
electron-impact ionization, recombination,
excitation, de-excitation, radiative recombination, spontaneous
decay, dielectronic recombination, autoionization, and electron
capture.  The atomic model for argon consisted of 16,000 levels
from all ionization stages with approximately 1500
over the lowest three ionization stages.
The non-LTE calculations also provide the
ionization fraction $f$ as a function of
$n_e$ and $T_e$\@.  The $f$ value is then used, in conjunction with
Eq.~(\ref{phase-shift-eq}) and the interferometer data, to provide an
independent  determination of $n_{tot} = n_i/f$.  If we assume that
$n_e= n_i$ ({\em i.e.}, all ions are singly ionized, a reasonable
assumption at our temperatures and densities), then $n_e=n_i=fn_{tot}$.

\begin{figure}[!htb]
\includegraphics[width=2.8truein]{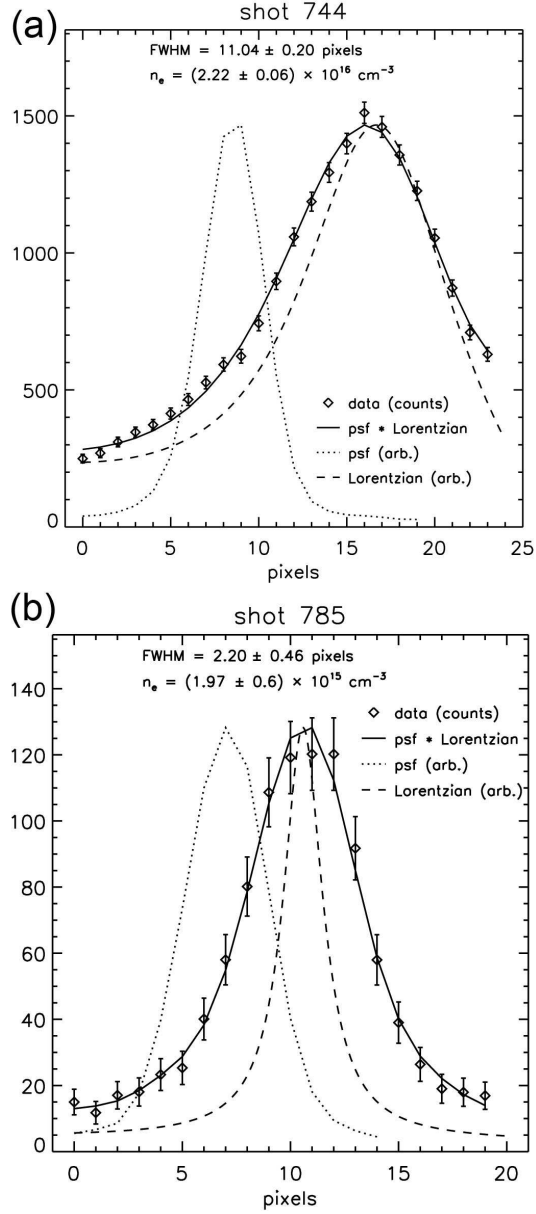}
\caption{\label{lorentzians}Determination of electron density 
$n_e$ via Stark
broadening of the H$_\beta$ line for shots (a)~744 ($t=17$~$\mu$s, {\em nozzle} view),
and (b)~785 ($t=28$~$\mu$s, {\em chord} view).
Shown are the experimental data (diamonds with
error bars = $\pm \sqrt{\rm counts}$),
an overlay of the measured instrumental broadening profile 
(dotted line, labeled as ``psf''
for point spread function), and a Lorentzian H$_\beta$ profile (dashed line)
that gives the best fit (minimum $\chi^2$) of
the convolution (solid line) of the psf and the Lorentzian to the data.
The $n_e$ is calculated from the Lorentzian full-width half-maximum (FWHM)
using Eq.~(\ref{stark-eq}),
and the uncertainty in $n_e$ is based on the standard deviation of
the FWHM determined from curve-fitting.}
\end{figure}

Stark broadening of the H$_\beta$ line is analyzed for the
{\em nozzle} and {\em chord} views (see Fig.~\ref{exp-setup}).
The latter coincides with the position of the $Z=41.4$~cm interferometer chord.
Figure~\ref{lorentzians} shows examples, from two shots ({\em nozzle} and 
{\em chord} views, respectively), of how $n_e$ is determined by
fitting the convolution of the measured instrumental broadening
(point spread function)
and a Lorentzian profile to the experimentally measured H$_\beta$ spectral feature.
Curve-fitting is performed using the IDL (Interactive Data Language,
\texttt{http://www.exelisvis.com/IDL}) routine 
\texttt{curvefit}.  The $n_e$ is determined via 
\begin{equation}
n_e~({\rm cm}^{-3})= 2.53 \times10^{14} \left[\frac{\rm FWHM~(nm)}
{\alpha_{1/2}}\right]^{3/2} = 1.50\times 10^{13} \left[\frac{\rm FWHM~(pixels)}
{\alpha_{1/2}}\right]^{3/2},
\label{stark-eq}
\end{equation}
where FWHM is the full-width half-maximum of the Lorentzian fit, and 
$\alpha_{1/2}$ (in our case $=0.085$) is the so-called reduced half-width that scales
as line shape and has been tabulated\cite{stehle99} for many hydrogen lines for
the temperature range 0.5--4~eV and density range 
$10^{14}$--$10^{18}$~cm$^{-3}$.
The value of 0.152~nm/pixel for our spectrometer system has been used
in Eq.~(\ref{stark-eq}).  Table~\ref{ne-summary} summarizes our $n_e$
results obtained by Stark broadening analysis, and their corresponding
viewing chords, times, and relative positions in the jet. 
The $n_e \approx 2\times 10^{16}$~cm$^{-3}$ near
the gun nozzle falls by approximately one order of magnitude after the jet propagates
approximately 41~cm.
Unfortunately, the number of shots on which we could successfully perform the 
Stark broadening analysis was limited because the H$_\beta$ line
typically is obscured by a stronger nearby argon line.
In the shots analyzed, the nearby argon line is subtracted out
prior to performing Stark broadening analysis on the H$_\beta$ line.

\begin{figure}[!htb]
\includegraphics[width=3.truein]{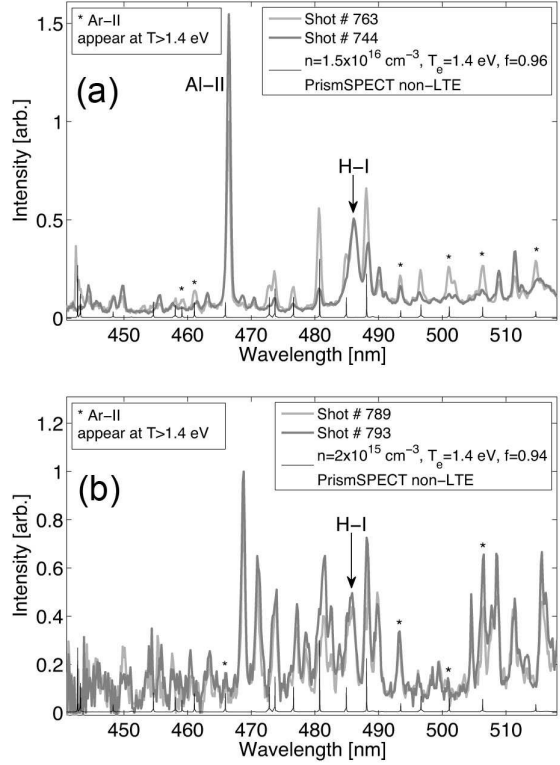}%
\caption{\label{spect-data}Intensity versus wavelength
from the survey spectrometer with
spectrometer views at the (a)~{\em nozzle} ($t=17$~$\mu$s)
and (b)~{\em chord} ($t=36$~$\mu$s) positions.
Shown in both plots are also the argon spectra of non-LTE PrismSPECT calculations 
with the parameters
indicated in the respective legends.  Appearance of the Ar~\textsc{ii} lines
in the data indicated by asterisks implies
that peak $T_e\ge 1.4$~eV according to the PrismSPECT calculations.}
\end{figure}

Next, we use the spectroscopy results to obtain an estimate of $T_e$.
Figure~\ref{spect-data}
shows spectrometer data for the (a)~{\em nozzle} position 
at $t=17$~$\mu$s (corresponding to the rising edge of the jet)
and (b)~{\em chord} position at $t=36$~$\mu$s (corresponding to the jet bulk).
The figure also shows calculated PrismSPECT non-LTE argon spectra for comparison.
The Ar~\textsc{ii} lines that appear in the experimental data
only appear for $T_e \ge 1.4$~eV
in the PrismSPECT non-LTE calculations, thus providing a lower
bound estimate of the peak $T_e$ in the jet.
The calculations show that $f=0.96$ (for $n_e=1.5\times10^{16}$~cm$^{-3}$ and
$T_e=1.4$~eV) and $f=0.94$ (for $n_e=2\times 10^{15}$~cm$^{-3}$ and $T_e=1.4$~eV)
for the {\em nozzle} and
{\em chord} positions, respectively.  
An upper bound on $T_e$
is also estimated based on the argon spectra centered around
794~nm (not shown).  An Ar~\textsc{ii} line visible in the calculations
at $n_e=2\times10^{16}$~cm$^{-3}$ and $T_e=1.7$~eV, but
not visible in our measurements (shots 569-572 and 655-706), indicates
an upper bound on the peak $T_e$ of 1.7~eV for the {\em nozzle} view.
This upper bound also justifies our assumption that all argon ions are singly ionized.
For the rest of the paper,
we use $T_e=1.4$~eV, which is a more appropriate jet-averaged
value, in estimating $T_e$-dependent quantities.  Although $T_i$ has not
been directly measured yet, it is reasonable for the purposes of this paper
to assume that $T_i\approx T_e$ because the ion-electron thermal 
equilibration time
$\bar{\nu}_{\epsilon}^{-1} = (3.2\times 10^{-9}Z_{eff}^2\ln\lambda n_e/
\mu T^{3/2})^{-1}$, estimated to be 0.2~$\mu$s, is very fast compared to the
jet evolution occurring over many tens of $\mu$s (where we have used
$Z_{eff}=1$, $\ln \lambda =4.74$, $n_e=2\times 10^{16}$~cm$^{-3}$, $\mu=m_i/m_p=40$,
and $T=1.4$~eV in estimating $\bar{\nu}_{\epsilon}$).

Finally, using the {\em chord} position $f=0.94$ determined above,
an estimated interferometer chord path length $\approx 10$~cm at
the $Z=41.4$~cm chord (see Sec.~\ref{jet-structure}),
and Eq.~(\ref{phase-shift-eq}), we obtain an independent determination
of $n_e=fn_{tot}$ from the interferometry data.
The results are shown in Fig.~\ref{int-spect-combo}, showing reasonable
agreement in $n_e$ obtained via spectroscopy and interferometry.
Table~\ref{jet-params} summarizes all the experimentally measured jet parameters
at both the {\em nozzle} and {\em chord} positions. 

\subsubsection{Remark on the jet magnetic field}
\label{mag-field}

The evolution of the jet magnetic field $B$ after the jet
exits the railgun has not been measured yet on PLX\@.
The gun bore magnetic probes show that the field strength within the gun bore
is on the order of several Tesla.    The classical diffusion time
of $B$ is approximately $\tau_D\sim \mu_o \delta^2/\eta_\perp$,
where $\mu_0=4\pi\times 10^7$~H/m
is the vacuum permeability, $\delta\approx 2$~cm the gradient scale length of $B$, 
and $\eta_\perp$ the perpendicular
Spitzer resistivity.  For $T_e=1.4$~eV, $n_e=2\times 10^{16}$~cm$^{-3}$,
$Z_{eff}=1$, and $\ln \lambda=4.74$, $\eta_\perp=2.95\times 10^{-4}$~$\Omega$m
and $\tau_D= 1.7$~$\mu$s.  Thus, for $V_{jet}\approx 3$~cm/$\mu$s, the magnetic
field will decay to $e^{-3}=0.05$ of its original value after 5.1~$\mu$s or
15.3~cm of propagation.
Assuming a value of 3~T in the gun bore, the decayed value after 5.1~$\mu$s
will be 0.15~T, and the ratio of jet magnetic energy density
$B^2/2\mu_0=9.0 \times 10^3$~J/m$^3$ to kinetic energy density 
$\rho V_{jet}^2/2=3.01\times 10^5$~J/m$^3$ (assuming an argon density of $10^{16}$~cm$^{-3}$
and $V_{jet}=30$~km/s) will be about 0.03.
Thus, it is reasonable to ignore (to leading order) the effects of
the magnetic field on jet evolution over 0.5~m.  We defer the direct measurement of
jet magnetic field evolution to future work.

\subsection{Jet structure and evolution}
\label{jet-structure}

In this sub-section, we present results on jet structure via
CCD image, photodiode, and interferometer data,
including
quantitative results on jet length inferred from photodiode and interferometer data,
and jet diameter inferred from CCD images and interferometer data.
Note that our jets have a primary leading structure with several
trailing structures, as seen in the
$Z=2.7$~cm trace after $t=40$~$\mu$s
in Fig.~\ref{pd-data}.  
By triggering a crowbar circuit to eliminate the ringing current
(results not shown in this paper), we have
successfully eliminated the trailing jet structures in both the photodiode and 
interferometer data.  In this paper, we focus only on the 
properties of the leading jet structure in non-crowbarred shots.

\subsubsection{Jet length and axial profile}
\label{jet-length}

\begin{figure}[!htb]
\includegraphics[width=2.5truein]{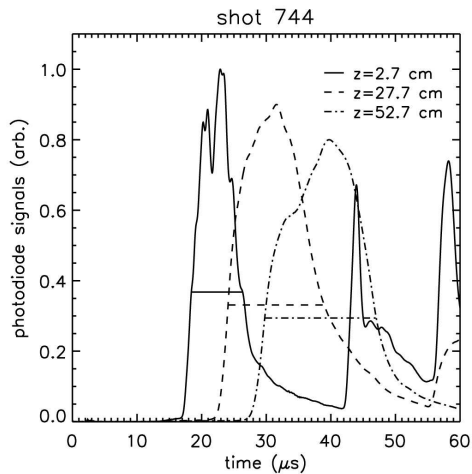}%
\caption{\label{velocity}Photodiode array signals versus time.  Horizontal
lines denote the duration $\Delta t_{jet}$ over which the signals are greater
than $1/e$ of the peak value (used in calculating
jet length and rate of axial expansion).  In this case, the velocities
are 28.7 and 30.5~km/s between the first and second pairs of photodiodes, respectively.
The $\Delta t_{jet}$ values are 7.9, 14.9, and 17.4~$\mu$s,
and the jet lengths are 22.8, 44.2, and 53.0~cm at $Z=2.7$, 27.7, 52.7~cm, respectively.}
\label{pd-data}
\end{figure}

We estimate the jet length $L$ using the full-width at $1/e$ of the maximum
of the photodiode (see Fig.~\ref{velocity} as an example)
and interferometer (see Fig.~\ref{744-int}(b) as an example) signals
versus time to get a $\Delta t_{jet}$ for each photodiode 
and interferometer signal. 
It follows that
$L\approx V_{jet}\Delta t_{jet}$, where $V_{jet}$ is determined
by the difference in arrival times of the signal peaks, as described in
Sec.~\ref{sec:velocity}.  The photodiodes
give $L_1=20.6\pm 3.5$~cm, $L_2=41.8\pm 4.4$~cm, and $L_3=48.9\pm 7.5$~cm
(where the variation is the standard deviation over the data set)
for the photodiode views at $Z=2.7$, 27.7, and 52.7~cm, respectively.
The velocities used 
are $V_{12}=28.9$~km/s, $(V_{12}+V_{23})/2=29.1$~km/s, and $V_{23}=29.4$~km/s 
(see Sec.~\ref{sec:velocity}) for calculating $L_1$, $L_2$, and $L_3$, respectively.
The interferometer chord at 41.4~cm, which is in between
the photodiode views at $Z=25$ and 50~cm, gives $L_{int}=46.7 \pm 6.5$~cm,
consistent with $L_2<L_{int}<L_3$, where the velocity used is $34.8$~km/s.

The rate of jet length expansion $\dot{L}$ is determined
from both the photodiode and interferometer data by taking $\Delta L/\Delta t$
between adjacent measurement positions.  From the photodiode data over the
entire data set, $\dot{L}_{12}= 24.4 \pm 5.4$~km/s and $\dot{L}_{23}=9.0 \pm 7.3$~km/s,
which correspond to the spatial ranges $Z=2.7$--27.7~cm and $Z=27.7$--52.7~cm,
respectively.  
From the interferometer data, $\dot{L}_{int} = 20.2 \pm 14.0$~km/s
for the spatial range $Z=35.0$--41.4~cm.  
If we assume $\dot{L}/2=2C_s/(\gamma-1)$
from hydrodynamic theory,\cite{landau87} where
$C_s=9.79\times10^5(\gamma Z_{eff} T_e/\mu)^{1/2}$, we get an independent
estimate of $T_e=1.8$~eV (for $\dot{L}=\dot{L}_{12}=24.4$~km/s,
$\mu=40$, $Z_{eff}=1$, and $\gamma=1.4$) that is in reasonable agreement with
the range of $T_e=1.4$--1.7~eV determined in Sec.~\ref{n-T-f}.

The jet axial profile can be inferred, qualitatively, from the photodiode
and interferometer signals versus time.  The time dimension can be converted
approximately to a spatial dimension by multiplying by $V_{jet}\sim 30$~km/s or
3~cm/$\mu$s (which was done above to obtain $L=V_{jet}\Delta t_{jet}$).
The qualitative jet axial profile as it comes out of 
the railgun is best observed in the $Z=2.7$~cm photodiode trace (Fig.~\ref{velocity}),
which shows a sharp rising edge lasting about 2~$\mu$s (6~cm).
This is followed by some structure over about 5~$\mu$s (15~cm)
and then by a falling tail that decays to about 10\% of the peak
intensity over about 10~$\mu$s (30~cm).  Figure~\ref{int-spect-combo}
gives a good representation of the jet axial profile farther downstream ($Z=41$~cm),
based on interferometer density data.

For an average $V_{jet}=V_{12}=28.9$~km/s (see
Sec.~\ref{sec:velocity}) as the jet exits the nozzle,
this corresponds to an average $\Delta t_{jet}\approx
L_1/V_{12}=7.1$~$\mu$s, which corresponds to the half-period of \Igun\ (see
Fig.~\ref{i-v}).  This
suggests that the jet length is inversely proportional to the frequency of the
pulsed-power railgun electrical circuit, and that the jet length can be
tailored via electrical circuit parameters.

\subsubsection{Jet diameter and radial profile}
\label{diameter}

The jet diameter $D$ is estimated by two methods.  The first method is examining
CCD image line-outs perpendicular to the jet propagation direction $Z$; this method
also provides information regarding the jet radial profile.  The second 
method uses the decay of the peak phase shift $\Delta \phi_{peak}$
of the eight interferometer chords
and the assumption of conservation of total jet mass to deduce the jet diameter.
We recognize that the two methods provide related but fundamentally
different information, {\em i.e.},
the first method relies on the intensity of emission from the jet plasma 
(which depends on both density and temperature) while
the second method derives from the interferometer phase shift (which depends
predominantly on the line-integrated
plasma density).  Nevertheless, the two methods give us meaningful 
quantitative estimates 
of the jet diameter and also provide qualitative
information on the shape of the radial profile.

Figure~\ref{jet-diameters}(a) shows an example of a CCD image vertical 
line-out at $Z=41.4$~cm,
and the estimate of $D$ based on the full-width at $1/e$ of the maximum.
The time of this CCD image ($t=36.0$~$\mu$s) corresponds approximately to when
the peak emission reaches $Z\approx 41$~cm (based on photodiode data).
The line-out is taken
after the CCD image is rotated slightly ($\approx 2^\circ$)
such that the jet propagation axis
is horizontal.  The horizontal-pixel value ($x$) 
of the CCD image is converted to the $Z$ coordinate (cm) based
on the formula $Z(x)=68.5 - 10^5x^2 - 0.041x$ (where $0<x<1023$).
In addition, the cm/pixel conversion $\Delta Y$ in the vertical direction of
the CCD image is obtained using the formula
$\Delta Y(x)=2\times 10^{-5}x+0.041$~cm/pixel.  The dependence of $\Delta Y$ on
$x$ is due to the camera perspective, {\em i.e.},
nearer objects appear larger in the image.
These formulae were determined from a CCD image of a meter stick held to the
end of the railgun nozzle.  
Figure~\ref{jet-diameters}(b) shows $D$ versus time at $Z=41.4$~cm determined
using the full-width at $1/e$ method from a series of CCD images (shots 780--819).
Note that $D$ determined via this method is fairly constant, with $D\sim 10$~cm
from $t=26$--38~$\mu$s, but then $D$ increases after 38~$\mu$s,
which corresponds
to the time when the peak of the leading jet structure is passing through
$Z=41$~cm (which can be deduced from Fig.~\ref{velocity}).
The expansion in $D$ at later times implies a lower $M$, suggesting
that the trailing part of the jet is either hotter, slower, or both.

We also consider CCD image vertical line-outs at different $Z$ positions
from a single shot at a single time ($t=30.0$~$\mu$s),
as shown in Fig.~\ref{jet-diameters}(c).  The diameters as determined by the 
full-width at $1/e$ of each peak are 8.2, 8.5, 8.5, 9.1, and 10.6~cm for
$Z=5$, 10, 15, 20, and 25~cm, respectively.  These emission line-outs provide
information on the radial profile of the jet, and
can provide further information on $n_e$ and $T_e$ profiles if compared
with synthetic emission profiles generated using spectral modeling
codes such as Spect3D.\cite{macfarlane07}

\begin{figure}[!htb]
\includegraphics[width=3truein]{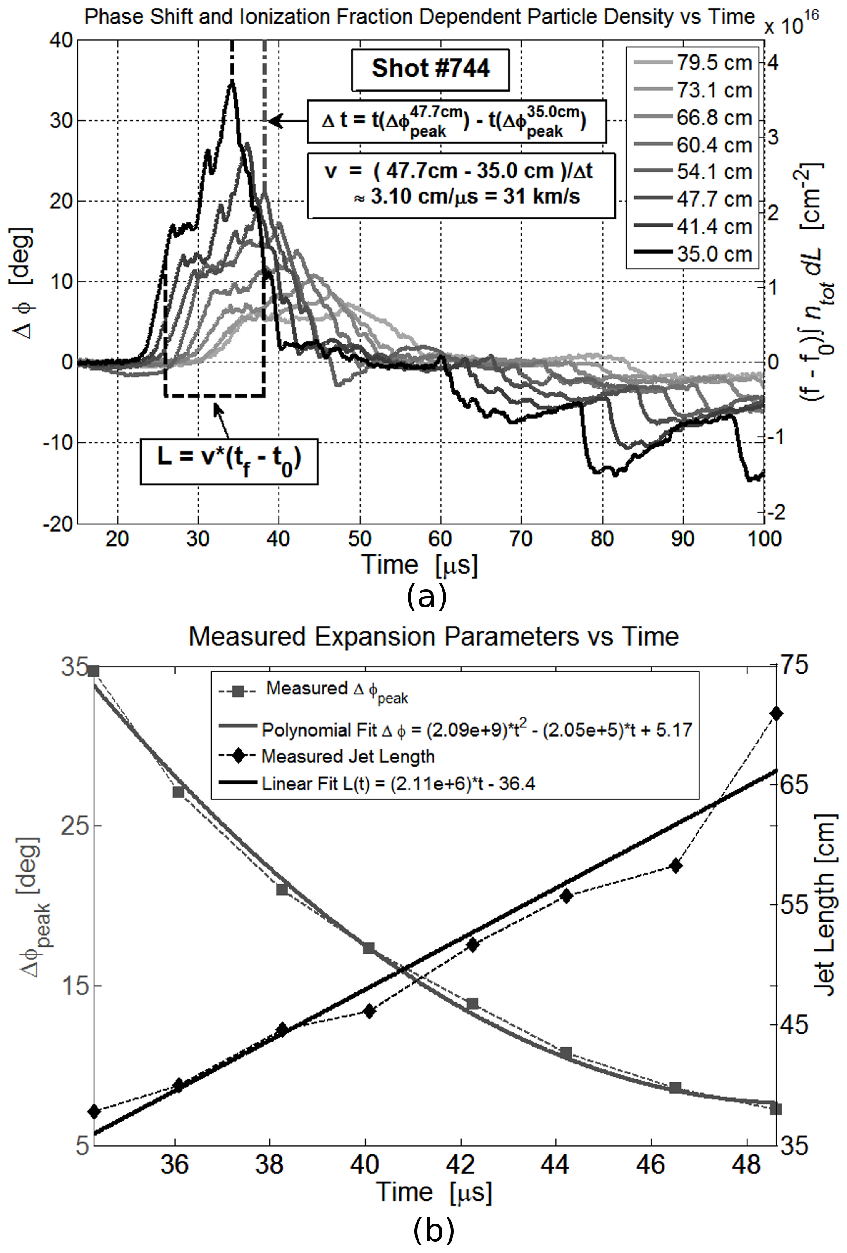}%
\caption{\label{744-int}(a)~Interferometer phase shifts versus time for all eight
interferometer chords (distances $Z$ from the jet nozzle are indicated in the legend). 
Methodologies for
calculating jet velocity and length are shown.  (b)~The peak
phase shift $\Delta \phi_{peak}$ (squares) and jet length (diamonds)
are shown for each chord
at the time at which $\Delta \phi_{peak}$ for each chord occurs.  Also shown
are analytic fits to each set of data.}
\end{figure}

\begin{figure}[!htb]
\includegraphics[width=2.5truein]{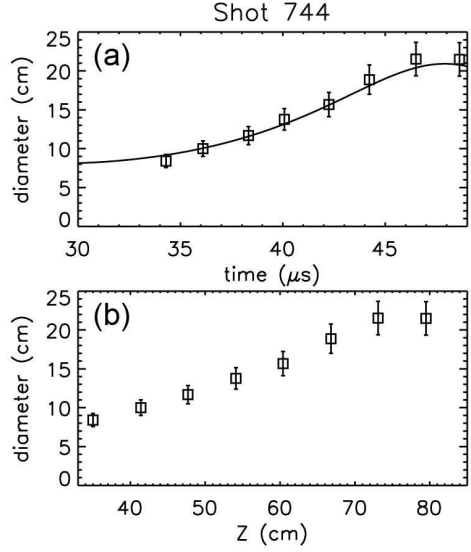}
\caption{\label{744-diameter-vs-time}Plasma jet diameter versus (a)~time
and (b)~$Z$, where the
squares are from the data at each interferometer chord, and the solid line is from 
Eq.~(\ref{D-from-phi-L}) using the
fitting functions for $\Delta \phi_{peak}$ and $L$ given in Fig.~\ref{744-int}(b).
Error bars represent $\pm 10$\% uncertainty in the determination of $L$, which
is used in Eq.~(\ref{D-from-phi-L}).}
\end{figure}

Next, we evaluate jet radial expansion using interferometer
data.  Figure~\ref{744-int}(a) shows that 
$\Delta \phi_{peak}$ decreases with increasing $Z$
chord position.  Figure~\ref{744-int}(b) shows $\Delta \phi_{peak}$ versus
time for each chord, and fits a quadratic function
in time to the data points.  From Eq.~(\ref{phase-shift-eq}), it is apparent that
$\Delta \phi_{peak} \sim n_{tot}D$ by assuming $\int n_{tot}{\rm d}l = n_{tot}D$
and constant $f$.
By invoking conservation of total jet mass, we obtain the 
relationship\cite{merritt12b}
\begin{equation}
n_{tot}(t)D(t)^2L(t) \sim \Delta \phi_{peak}(t)D(t)L(t)={\rm constant} 
\Rightarrow D(t) \propto \frac{1}{\Delta\phi_{peak}(t)L(t)},
\label{D-from-phi-L}
\end{equation}
giving $D(t)$ in terms of experimentally measured quantities $\Delta \phi_{peak}(t)$
and $L(t)$, which are both shown in Fig.~\ref{744-int}(b) for shot~744 as an example.
By using the analytic fits to $\Delta \phi_{peak}(t)$ and $L(t)$
given in the legend of Fig.~\ref{744-int}(b), it is
straightforward to calculate $D(t)$ to within a constant that can be determined
using $D(t=36~\mu{\rm s})\approx 10$~cm from Fig.~\ref{jet-diameters}(b).
The result is shown in Fig.~\ref{744-diameter-vs-time}.
Note that the diameter obtained at $Z\approx 41$~cm obtained from CCD image line-outs
is likely an underestimate
because the jet emission falls off more sharply than density
if there is a peaked temperature profile.  We regard the uncertainty of our
reported diameter results based on CCD line-outs to be around a factor of two.
The nominal radial expansion rate is $\approx 7$~km/s (between 36 and 43~$\mu$s),
which is within the range $\dot{L}/2=10.1\pm 7.0$~km/s obtained
between the interferometer chords at 35.0 and 41.4~cm.

Finally, we show examples of the plasma jet
radial density profiles deduced using Abel inversion\cite{hutchinson02}
of two chord-arrays of interferometer data from
a single shot (1106).  This shot is not part of the data set analyzed in the rest
of this paper because the Abel inversion analysis requires that the interferometer
chords be arranged with different impact factors relative to
the jet propagation axis $Z$, as shown in Fig.~\ref{abel}(a).
For the main data set in this paper, the interferometer
chords were arranged to intersect the $Z$ axis at different values of $Z$\@.
Our Abel inversion analysis assumes cylindrical symmetry for the jet and
four radially concentric zones of uniform density in each zone.  The radii of the
center of the zones correspond to the four interferometer chord positions
(impact factors), respectively, for each 
$Z$ location, {\em i.e.}, 5, 10, 15, and 20~cm for the $Z\approx 60$~cm chord array.
The results of the Abel inversion are shown in Figs.~\ref{abel}(b) and \ref{abel}(c),
which show $(f-f_0)n_{tot}$ and radial line-outs
of $(f-f_0)n_{tot}$ and $n_e$, respectively.  From Fig.~\ref{abel}(c), it can
be seen that the jet density profile is about a factor of two larger than
the density profile from CCD image line-outs shown in Figs.~\ref{jet-diameters}(a)
and \ref{jet-diameters}(c).

\section{Conclusions and summary}
\label{conclusions}

The issues we address in this work (Sec.~\ref{issues}) are determining the jet parameters,
evaluating the evolution of the jet parameters as the jet propagates over about
0.5~m, and characterizing the jet axial and radial profiles to the extent possible.
These issues are important for accurate assessments
of the formation of imploding spherical
plasma liners using merging plasma jets.  The experimental data, while not
in all cases definitive due to diagnostic limitations, offer
crucial information about jet initial conditions and constraints 
on jet evolution that can enhance the accuracy of numerical modeling 
predictions.\cite{loverich10jfe,thoma11pop,cassibry12,cassibry12b}
We reiterate that our plasma jet diagnostic measurements are
all averaged over their viewing-chords, and therefore our reported
results are mostly chord-averaged quantities.  

We have presented results
on jet parameters, summarized in Tables~\ref{ne-summary} and \ref{jet-params},
showing that we are within a factor of 2--5
of what is needed to field the thirty-jet imploding
plasma liner formation experiments specified in the PLX design.\cite{hsu12,cassibry12b}
Experiments at HyperV Technologies, using the same railgun design as the railgun
used in this work but operating at $\gtrsim 500$~kA,
have demonstrated the simultaneous achievement of the full
PLX design parameters:  density $\approx 10^{17}$~cm$^{-3}$,
velocity $\approx 50$~km/s, and jet mass $\approx 8$~mg.  Those results will
be reported elsewhere.
We operated at reduced railgun current ($\lesssim 300$~kA)
in order to extend the railgun lifetime and maximize
the amount of data collected. Although not emphasized in this paper, we have also
learned how to operate the railgun reliably and taken note of where technological
improvements should be made (mostly in the details of the
pulsed-power system) for a thirty-gun experiment.

The need to understand the
evolution of the jet as it travels a distance of about 0.5~m 
was a primary motivation for this work.  To this end, our diagnostics were focused on
two locations:  the {\em nozzle} position right at the nozzle exit and the {\em chord}
position about $Z=41$~cm away from the nozzle.  The diagnostic views were 
transverse or mostly transverse to the jet propagation axis.
We were able to determine that average $n_e$ fell by about one order of magnitude
(from $2\times 10^{16}$ to $2\times 10^{15}$~cm$^{-3}$) over this distance,
while average $T_e$ and $f$ remained nearly constant 
(within our measurement resolution).  
We were also able to determine the approximate jet length and diameter 
(see Table~\ref{jet-params}), and their
evolution, showing that the jet volume increased by about one order of magnitude
which is consistent with the drop in $n_e$.  As discussed in Sec.~\ref{issues},
the purely hydrodynamic prediction of constant-$M$ jet expansion has a large uncertainty
due to a lack of precise knowledge of the exact expansion rate.  This
uncertainty is further
exacerbated by the theory not accounting for radiative cooling which is
expected to be important in our parameter regime.
Using our measured jet parameters, the hydrodynamic theory predicts
that the jet volume could increase by up to a factor of approximately 150.
Our experimentally observed volume increase and density drop by about
a factor of ten provides a useful constraint
for validating numerical modeling results (that also
include the effects of atomic physics) on jet propagation.

Adiabatic expansion of the jet would dictate
that $nTV^\gamma={\rm constant}$.  With a density drop of ten, a volume increase
of ten, and $\gamma=1.4$, the average temperature should have dropped by a factor of
about 2.5.  The discrepancy between the latter and what was observed
could potentially be explained by Ohmic heating associated with magnetic energy
dissipation.  The initial plasma jet thermal $\beta$ is $< 0.01$, and thus the
magnetic energy dissipation (discussed in Sec.~\ref{mag-field}) could easily
balance thermal and radiative losses during jet propagation.

\begin{figure}[!htb]
\includegraphics[width=3.truein]{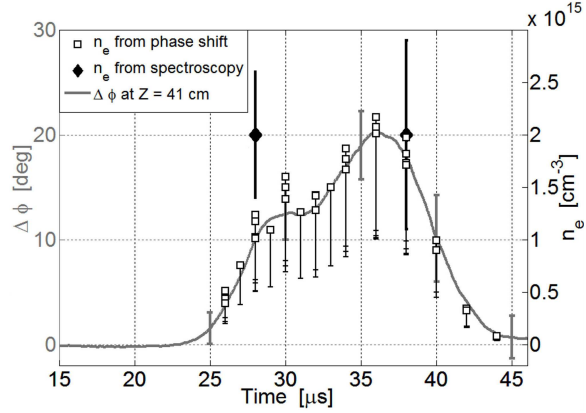}
\caption{\label{int-spect-combo}Interferometer phase shift $\Delta \phi$
averaged over shots 775--819 for the $Z=41.4$~cm chord (left hand y-axis)
and electron density $n_e$ (right hand y-axis) versus time.  The square data points
are derived
from Eq.~(\ref{phase-shift-eq}) using $f=0.94$, interferometer
$\Delta \phi$ data from the $Z=41.4$~cm chord, and the corresponding path length estimate
based on the jet diameter $D$ obtained from CCD line-outs (discussed in
Sec.~\ref{diameter}).  Error bars on the square data points
represent uncertainty based on doubling the path length, which would halve the
average $n_e$.  The two diamond data points are 
from Stark broadening
analysis of the H$_\beta$ line from spectroscopy
(shots 785 and 790).  Each discrete data point corresponds
to a separate shot.}
\end{figure}

\begin{figure}[!htb]
\includegraphics[width=2.5truein]{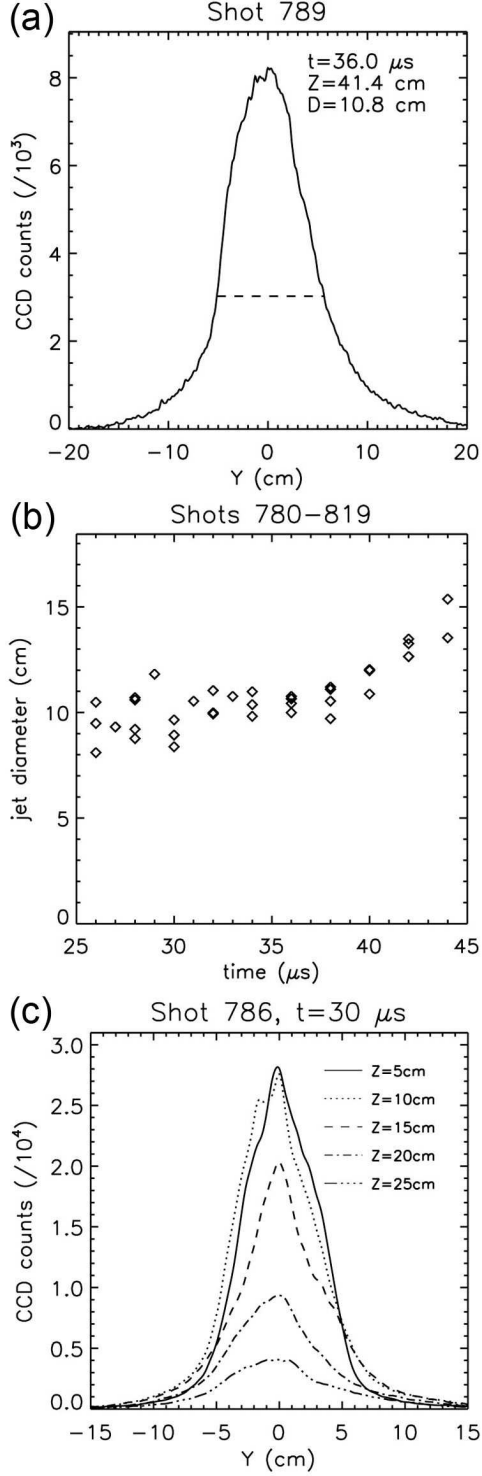}
\caption{\label{jet-diameters}
(a)~CCD image line-out versus $Y$ (perpendicular to jet propagation direction)
and the full-width (FW) at $1/e$ (dashed line) used to define the jet diameter $D$;
(b)~$D$ versus time at $Z=41.4$~cm
from CCD line-outs (shots 780--819); (c)~CCD
line-outs at different $Z$ positions from a single shot, with FW at $1/e$
jet diameters
of 8.2, 8.5, 8.5, 9.1, 10.6~cm for the increasing $Z$ values, respectively.}
\end{figure}

\begin{figure}[!htb]
\includegraphics[width=3.truein]{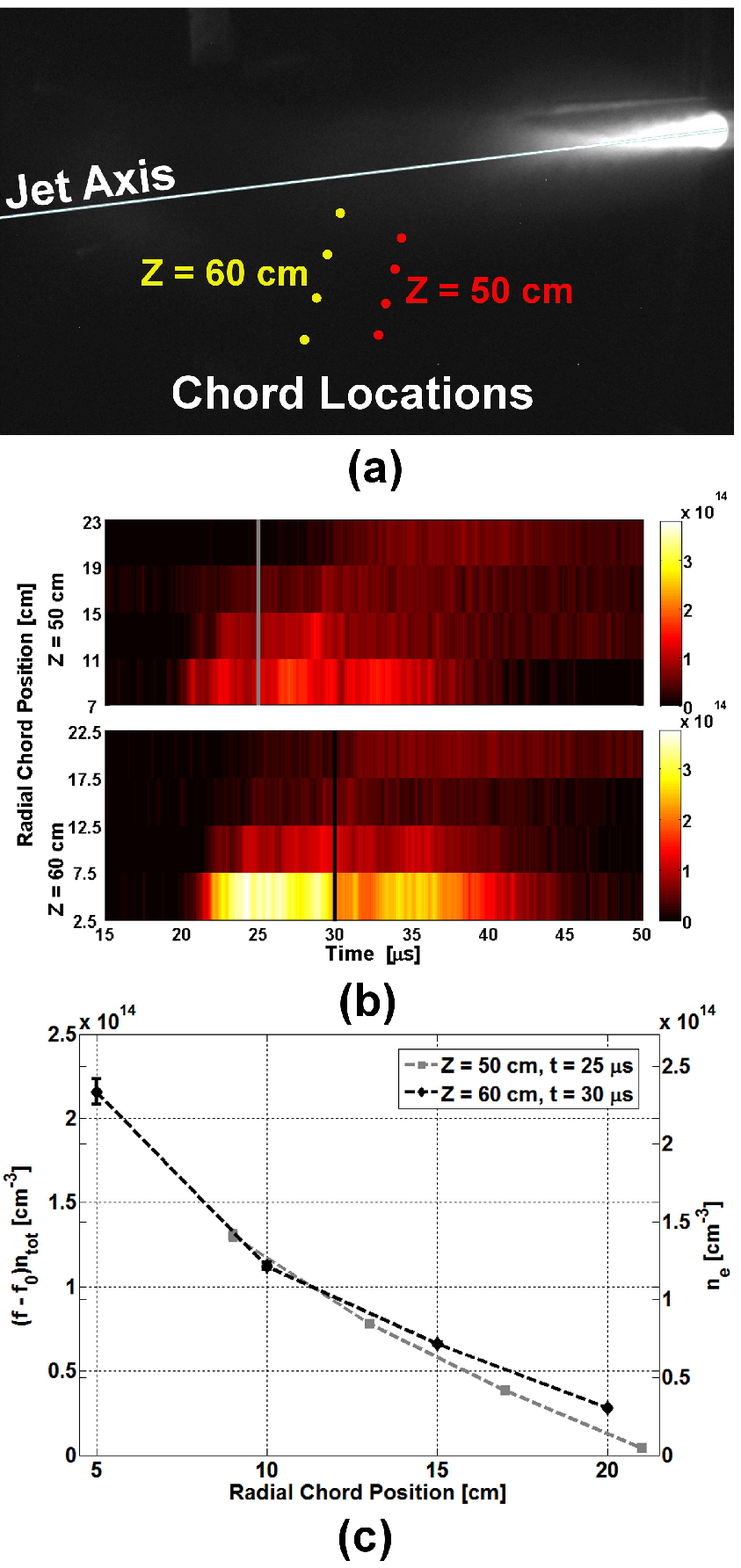}
\caption{\label{abel}
(a)~Location of interferometer chords overlayed on the CCD image for shot 1106,
on which we performed an Abel inversion analysis 
to obtain jet radial density
profiles (for the yellow and red chord arrays separately).  (b)~Plots of 
$(f-f_0) n_{tot}$ [cm$^{-3}$]
versus radial chord position (relative to the jet axis)
and time for the $Z\approx50$~cm (top) and 
$Z\approx 60$~cm  (bottom) chord positions; vertical lines indicate
the times for which line-outs are shown next.  (c)~$(f-f_0)n_{tot}$ 
and $n_e$ (data points) versus radial chord position at different times; 
$f=0.94$ 
and a maximum path length of 44~cm (corresponding to the
path length of the innermost chord at $Z\approx 60$~cm)
were used in calculating $n_e$.  Error bars represent
uncertainty in the chord positions of $\pm 0.5$~cm and
also account for the slight variation in $Z$ of each chord array.}
\end{figure}

The radial and axial profiles of the jet are important variables for determining
the dynamics of subsequent jet merging.  The profiles are also needed for
accurate numerical modeling of jet merging and plasma liner formation.
We have assessed the radial and axial profiles to the extent possible
via direct experimental measurements, and they are shown in 
Figs.~\ref{int-spect-combo}, \ref{jet-diameters}(a), 
\ref{jet-diameters}(c), and \ref{abel}(c).  The 
profiles inferred from CCD image line-outs are dependent on both $n_{tot}$ and $T_e$,
whereas the profiles inferred from interferometer data are dependent
mostly on $n_{tot}$.  If $T_e$ is peaked in both the
radial and axial directions, as can be reasonably expected, then this means
that the CCD line-outs will underestimate the width of the density profile.
These profile data provide the opportunity to validate numerical modeling
results by comparing the experimental profiles with synthetic data from post-processed
numerical simulation results.

In summary, we have reported experimental results on the parameters, structure, and
evolution of high-$M$ argon plasma jets launched by a pulsed-power-driven
railgun.  
An array of thirty such jets has been proposed as a way to form imploding
spherical plasma liners to reach 0.1--1~Mbar of stagnation pressure.  Imploding
plasma liners have potential applications as a standoff driver for MIF
and for forming repetitive cm-, $\mu$s-, and Mbar-scale plasmas for HED scientific
studies.


%
%

%

\begin{acknowledgments}
We thank R. Aragonez, D. Begay, D. Hanna, S. Fuelling, 
D. Martens, J. Schwartz, D. van Doren (also for
Figs.~\ref{plx-schematic} and \ref{gun-schematics}), and W. Waganaar
for their contributions toward PLX facility and diagnostic design/construction;
our colleagues at FAR-TECH, Prism Computational Sciences,
Tech-X, and Voss Scientific for discussions
and collaboration on jet theory and modeling;
T. Intrator and G. Wurden for loaning numerous items of laboratory equipment;
and Y. C. F. Thio for encouragement and extensive discussions.
This work was sponsored
by the Office of Fusion Energy Sciences of the U.S. Department
of Energy.
\end{acknowledgments}


%

\newpage
\begin{table}[!h]
\begin{tabular}{lcccc}
\hline
\hline
shot & view & time ($\mu$s) & position in jet & $n_e$~cm$^{-3}$ \\
\hline
744 & nozzle & 17 & rising edge & $2.2\pm 0.1 \times 10^{16}$\\
763 & nozzle & 17 & rising edge & $1.6\pm 0.3 \times 10^{16}$\\
738 & nozzle & 30 & late in jet & $8.6 \pm 0.0 \times 10^{15}$\\
771 & nozzle & 35 & late in jet & $5.8 \pm 0.1 \times 10^{15}$\\
785 & chord & 28 & rising edge & $2.0\pm 0.6 \times 10^{15}$\\
790 & chord & 38 & just past peak & $2.0\pm 0.9 \times 10^{15}$\\

\hline
\hline
\end{tabular}  
\caption{Summary of sight-line-averaged
electron density determined via Stark broadening analysis of
the H$_\beta$ spectral line (486.1~nm).
{\em Nozzle} view corresponds to just outside the railgun
nozzle, and {\em chord} view corresponds to $Z\approx 41$~cm (see Fig.~\ref{exp-setup}).
The {\em chord} position results are also
shown graphically in Fig.~\ref{int-spect-combo}.  The quoted
uncertainties in $n_e$ are the standard deviations obtained from
curve-fitting.}
\label{ne-summary}
\end{table}

\begin{table}[!h]
\begin{tabular}{lcc}
\hline
\hline
 & nozzle ($Z\approx 2$~cm) & chord ($Z\approx 41$~cm)\\
\hline
$n_e$ (cm$^{-3}$) & $2\times 10^{16}$ & $2\times 10^{15}$\\
$T_e$ (eV) & 1.4 & 1.4 \\
$V$ (km/s) & 30 & 30\\
$f$ & 0.96 & 0.94\\
$L$ (cm) & 20 & 45\\
$D$ (cm) & 5 & 10--20\\
\hline
\hline
\end{tabular}  
\caption{Summary of plasma jet parameters at the {\em nozzle} and {\em chord}
positions, as determined by experimental measurements.  The $n_e$
and $T_e$ values represent jet-averaged values.}
\label{jet-params}
\end{table}

\begin{figure}[!htb]
\includegraphics[width=3.truein]{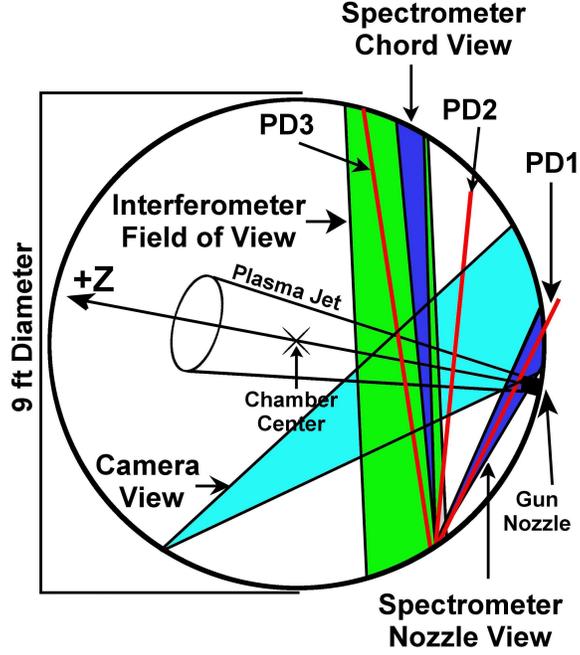}%
\caption{\label{exp-setup}Diagnostic setup for the experiments
reported in this paper.
The spectrometer {\em chord} view corresponds to the $Z=41.4$~cm interferometer chord,
and the spectrometer {\em nozzle} view is at the exit of the railgun nozzle.
PD1, PD2, and PD3 are the three photodiode views.}

\end{figure}

\begin{figure}[!htb]
\includegraphics[width=6.5truein]{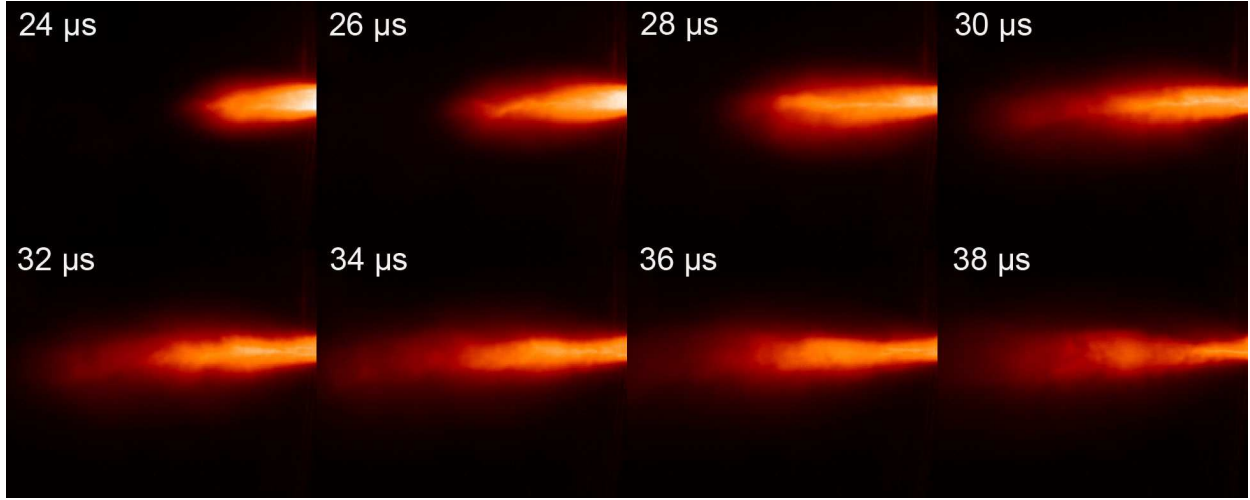}%
\caption{\label{images}Plasma jet evolution as recorded by the CCD camera over 8
separate shots (800 and 784--790).  The railgun nozzle is at the very
right edge of each image.  The images show the logarithm of the CCD intensity 
in false color.}
\end{figure}

\end{document}